\documentstyle[twocolumn,prc,aps,epsfig]{revtex}
\newcommand{\be}{\begin{eqnarray}}
\newcommand{\ee}{\end{eqnarray}}

 \newcommand{\gsim}{\mathrel{\hbox{\rlap{\lower.55ex \hbox {$\sim$}}
                   \kern-.3em \raise.4ex \hbox{$>$}}}}
\newcommand{\lsim}{\mathrel{\hbox{\rlap{\lower.55ex \hbox {$\sim$}}
                   \kern-.3em \raise.4ex \hbox{$<$}}}}

\def\roughly#1{\mathrel{\raise.3ex\hbox{$#1$\kern-.75em%
\lower1ex\hbox{$\sim$}}}}
\def\lsim{\roughly<}
\def\gsim{\roughly>}

\newcommand{\eq}{\begin{equation}}
\newcommand{\eqx}{\end{equation}}
\newcommand{\eqn}{\begin{eqnarray}}
\newcommand{\eqnx}{\end{eqnarray}}

\begin{document}

\twocolumn[\hsize\textwidth\columnwidth\hsize\csname @twocolumnfalse\endcsname

\title {Prompt  Quark Production by exploding Sphalerons}
\author {Edward Shuryak and Ismail Zahed}
\address {
     Department of Physics and Astronomy, State University of New York,
     Stony Brook, NY 11794
}

\date{\today}
\maketitle
\begin{abstract}
Following recent works on production and subsequent explosive decay of QCD
sphaleron-like clusters,  
we discuss the mechanism of quark pair production in this process. 
We first show how the gauge field explosive solution of Luscher and
Schechter can be achieved by non-central
conformal mapping from the O(4)-symmetric solution. Our main result
is a new solution to the Dirac equation in real time in this configuration,
obtained by the same inversion of the fermion O(4) zero mode.
It explicitly shows how the quark acceleration occurs, 
 starting from the spherically O(3) symmetric zero energy chiral quark
state to the final spectrum of non-zero energies.
 The sphaleron-like clusters with any Chern-Simons number always
produce ${\rm N_F}\,{\overline {\bf L}}{\bf R}$  quarks, 
and the antisphaleron-like clusters the chirality opposite. 
 The result are 
relevant for hadron-hadron and nucleus-nucleus collisions at
large $\sqrt{s}$, wherein such clusters can be produced.
\end{abstract}
\vspace{0.1in}

]
\newpage

\section{Introduction}\label{intro}

Although it would hardly ever be possible to observe instanton-induced events in electroweak
theory, intense studies of such processes have started in this
context \cite{electroweak}.  Even prior to these studies, the decay of
the sphaleron solution \cite{Manton} has been studied in great details
numerically~\cite{sphaleron_decay}. It was found that the Chern-Simons
number at the exit state is frozen at some non-integer value. The question
then is what happens to the relationship between the anomaly and the
fermion chirality flow? Naively, the two concepts are related,
yet the fermion spectral flow can only accomodate integer numbers.
Should there be a generalization of index theorem for non-zero fields
radiated to infinity?
This question remained unanswered in general, although several points
of view
and partial results 
have been expressed in the litterature. In this paper we would not
address these questions in general, 
working out the solution
explicitly, by finding exact solutions to the Dirac equation in
exploding background fields, which demonstrates in detail how quark
production and subsequent acceleration occurs.

Recently the whole discussion of such questions has shifted into the QCD domain.
In refs~\cite{sz01,KKL} it was suggested that 
nonperturbative configurations composed of instanton/antiinstanton
play an important role in parton-parton scattering amplitudes at
high-energy, and may account for some part of the soft pomeron: the
estimates resulted in the slope and
intercept in approximate agreement with data. 
Detailed discussion of a specific gluonic clusters which are produced
have been made\cite{OCS}, with some proposed applications to high
energy heavy ion collisions \cite{ES_how}.  

  In the calculation of the cross section both the action of the gauge 
field corresponding to emitted gluons, or instanton-instanton
interaction, can be  naturally combined with the overlap factor with
the incoming high energy partons in the so called Landau method, in which
the semiclassical approximation with singular gauge configurations
is used~\cite{khlebnikov,DP94}. Pair singularities of the gauge field on the Euclidean
time axis play the role of coordinate infinity in the original Landau 
work on quantum mechanics. Those singularities are
key to interpolating between a vacuum configuration with zero
energy, and the escape configuration with finite energy $Q(T)$, where
$T$ measures the finite tunneling time. In a recent paper~\cite{jsz02} 
we have shown that the gluon multiplicity associated to the escaping
configurations at finite $Q$ follows from the configuration at the
sphaleron point $M_S$ through a pertinent rescaling of the sphaleron
size and energy density. 

In the present work we continue along the same line, focusing now on
the quark 
production  enforced by such exploding sphalerons configurations.
In section 2,  we show how a conformal mapping from the O(4)
conformally symmetric solutions of the Euclidean Yang-Mills
equations leads to the O(3) symmetric solutions  by
Luscher and Schechter~\cite{LS} which describes
the sphaleron explosion \cite{OCS}. In section 3, we construct the
normalizable zero mode solutions to the O(4) Yang-Mills
background, which includes the instanton zero modes as a
special case. In the central sections 4,5 we show that the solution to 
the massless Dirac equation  in the Minkowski background field of
exploding sphalerons can also be obtained by
the same conformal mapping, from the O(4) Euclidean zero
modes. We explicitly construct these  states and show that at the initial
Minkowski time t=0 those are zero energy states, while at
 asymptotically large time they reduce to a free quark
or free antiquark of specific chirality. We also calculate the
spectrum of the produced fermions.
In section 6,
we argue that pair production in the sphaleron (antisphaleron)
background is through pair level crossing at $t=0$. 
Our conclusions are in section 7.

\section{LS solution from O(4) by Inversion}

In this section we show that the spherically symmetric  O(3) Luscher-Schechter (LS) 
solution~\cite{LS} describing the explosion of the sphaleron-type clusters 
\cite{OCS} can be obtained  from the O(4) symmetric solution
through a pertinent inversion that breaks O(4). Similar considerations
for the $\phi^4$ model were originally discussed in~\cite{phi4}. In our
case, consider the O(4)  symmetric ansatz for the SU(2) gauge
configuration 

\be
A_{a\mu} (y) =\frac 1g\,\eta_{a\mu\nu}\,\partial_\nu F(y)
\label{Z1}
\ee
where

\be
F(y)=2\int_0^{\xi(y)} d\xi'\,f(\xi')\,\,,
\label{Z2}
\ee
with $\xi(y)={\rm ln} (y/\rho)$. The O(4) ansatz solves the Yang-Mills
equations provided that

\be
\frac{d^2f}{d\xi^2} = 4(f^2-1) (2f-1)\,\,.
\label{Z3}
\ee
The solution with the turning point at $\xi=0$ are expressible in terms
of elliptic sine and cosine,

\be
f (\xi) =\frac 12\left(1-\sqrt{1+\sqrt{2\epsilon}}\,
\,{\rm dn}\,\left(\sqrt{1+\sqrt{2\epsilon}}\,\xi\,; \frac 1m\right)\right)\,\,,
\label{Z4}
\ee
for $2\epsilon < 1$, while for $2\epsilon\geq 1$

\be
f (\xi) =\frac 12\left(1-\sqrt{1+\sqrt{2\epsilon}}\,
\,{\rm cn}\,\left({}^4\sqrt{8\epsilon}\,\xi\,; m\right)\right)\,\,,
\label{Z4x}
\ee
with $Q/M_S=2\epsilon$ and
$2m=1+1/\sqrt{2\epsilon}$. For $Q=0$ we recover the antiinstanton
solution, while for $Q=M_S$ we have the antisphaleron solution with

\be
f(\xi) = -\frac 12 \left( \frac{\sqrt{2}}{{\rm cos}
(\sqrt{2}\,\xi)}-1\right)\,\,.
\label{Z4}
\ee
The instanton and sphaleron configurations follow from a similar 
construction with a dual background $\eta\rightarrow\overline{\eta}$.

The O(4) solutions to the Yang-Mills equations with a turning point
at $\xi=0$ relates to the LS solution~\cite{LS}
with a turning point at $t=0$ by the conformal transformation

\be
(x+a)_\mu =\frac {2\rho^2}{(y+a)^2}\,(y+a)_\mu
\label{Z5}
\ee
with $a=(\rho,\vec{0})$, which maps the sphere $y^2=\rho^2$ onto
the upper-half of the x-space. Indeed, under the conformal
transformation (\ref{Z5}), (\ref{Z1}) transforms as

\be
A_{a\mu} (x) =&& \frac{(y+a)^2}{2\rho^2}\,\nonumber\\
&&\times\,
\left(g_{\mu\nu}-2\,\frac{(y+a)_\mu(y+a)_\nu}{(y+a)^2}\right)\,A_{a\nu}
(y)\,\,,
\label{Z6}
\ee
with $y$ solving (\ref{Z5}).
The LS solution in Minkowski space follows by an analytical continuation
$x_4\rightarrow it$, 

\be
&&A_0^a =-\frac{f(\xi )}{g}\,
\frac{8\rho\,tr\,n_a}{((t-i\rho)^2-r^2)((t+i\rho)^2-r^2)}\nonumber\\
&&A_i^a =+\frac{f(\xi )}{g}\,\nonumber\\
&&\times\,\frac{4\rho(\delta_{ai} (t^2-r^2+\rho^2)
+2\rho\,\epsilon_{aij}\,n_j+2n_a\,n_j)}
{((t-i\rho)^2-r^2)((t+i\rho)^2-r^2)}\,\,.
\label{Z6x}
\ee
The gauge configuration at $t=0$ satisfies $A_0^a=0$ and $\dot{A}_i^a=0$ 
which corresponds to zero chromoelectric field (zero momentum) and finite
chromomagnetic field. The $t=0$ point is a classical turning (or escape) point for
the gauge configuration, where Euclidean and Minkowski parts of the path
join together, see more on that in~\cite{OCS}.

\section{The O(4) Symmetric Zero Modes}

In this section we will derive the generic zero mode solution to the Dirac equation 
in the  O(4) background (\ref{Z1}). Our discussion in this section is
limited to just one massless quark flavor, and will paralell the
discussion in~\cite{book} for the instanton. For that, we introduce the conventions

\be
\sigma_\mu^\pm = ({\bf 1} , \mp i\,\vec{\sigma})
\label{Z7}
\ee
which are related to the t'Hooft symbols through\footnote{Here
$\eta^*_{a\mu\nu}=-{\overline{\eta}}_{a\mu\nu}$.}

\be
\sigma_\mu^-\sigma_\nu^+ =&&{\bf 1}_{\mu\nu}+i\,\eta_{a\mu\nu}\,\sigma_a=
\eta_{\alpha\mu\nu}\sigma_\alpha^+\nonumber\\
\sigma_\mu^+\sigma_\nu^- =&&{\bf 1}_{\mu\nu}-i\,\eta^*_{a\mu\nu}\,\sigma_a=
\eta^*_{\alpha\mu\nu}\sigma_\alpha^-\,\,.
\label{Z8}
\ee
In terms of (\ref{Z7}-\ref{Z8}) the Dirac operator in the 
background (\ref{Z1}) reads

\be
{\bf D}_-= \left(\partial_\mu +\frac 12\partial_\nu F\,\sigma_{\mu
c}^+\sigma_{\nu c}^-\right)\sigma_{\mu s}^+\,\,,
\label{Z9}
\ee
while in the dual background it reads

\be
{\bf D}_+= \left(\partial_\mu +\frac 12 \partial_\nu F\,\sigma_{\mu
c}^-\sigma_{\nu c}^+\right)\sigma_{\mu s}^-\,\,,
\label{Z10}
\ee
where the subsrcipts $c,s$ refer to the color, spin SU(2) matrices.
The coupling $g$ drops in the spectrum of ${\bf D}_\pm$ making the
quark states of order $g^0$, while the gauge fields are of order
$1/g$. In the strict semiclassical approximation, the quark effects
on the semiclassical gauge configurations can be ignored. 
We note that the combination

\be
{\bf P}_+ =
\frac 14 \sigma^-_{\mu s}\sigma^-_{\mu c} =
\frac 14 \sigma^+_{\mu s}\sigma^+_{\mu c}
\label{Z11}
\ee
projects onto a color-spin singlet. In particular $(U_+=\sigma_2)_{a\alpha}$
is an eigenstate of ${\bf P}_+$, i.e. ${\bf P}_+U_+=U_+$. Finally,
let $\pm$ {\bf L,R} chiralities for which
$\gamma_5{\bf \Psi}_\pm = \pm {\bf \Psi}_\pm$. In the $\eta$-field (${\bf
D}_+$) or $\overline{\eta}$-field  (${\bf D}_-$) the zero modes carry
specific chirality ${\bf D}_\pm {\bf \Psi}_\pm=0$.

To construct the normalizable
zero mode state in the configuration (\ref{Z1}),
we note that only the positive chirality state is `bound'

\be
\left(\sigma_{\mu s}^+\partial_\mu+2{\bf P}_+\,\sigma^-_{\mu
c}\partial_\mu\,F\right)\,{\bf \Psi}_+ =0
\label{Z13}
\ee
while the negative chirality state remains free. The general
solution to (\ref{Z13}) can be sought in the form

\be
{\bf \Psi}_+ (y) = {\bf C}\,\sigma^-_{\mu s}(\partial_\mu \Phi )\,U_+
\label{Z14}
\ee
up to an overall rigid color rotation and normalization (see below). Inserting (\ref{Z14}) into
(\ref{Z13}) and using the identity

\be
{\bf P}_+\,\sigma^-_{\alpha c}\sigma^-_{\beta s}
\,U_+=\delta_{\alpha\beta}\,U_+\,\,,
\label{Z15}
\ee
it follows that

\be
\Box\,\Phi= -2\,(\partial_\mu\,F)(\partial_\mu\,\Phi)
\label{Z16}
\ee
which is solved by

\be
\frac{d\Phi}{dy} = \frac 1{y^3}\,e^{-2\,F(\xi (y))}\,\,
\label{Z17}
\ee
where $F(\xi)$ is given by (\ref{Z2}). Using the identity (\ref{Z15}) 
for the norm of (\ref{Z14}) it follows that the overall
constant is fixed by

\be
{\bf C} = \left(2\int d^4y\,{\Phi'}^2 (y) \right)^{-1/2}\,\,.
\label{Z18}
\ee
The expression (\ref{Z14}) thus defines the Euclidean
normalized positive chirality (${\bf R}$) quark state in the 
O(4) background (\ref{Z1}). 
Since the O(4) Dirac spectrum is charge self-conjugate, the O(4)
background admits also a normalized negative chirality ({\bf L})
anti-quark state in the O(4) background, i.e. ${\bf \Psi}_-^\dagger$.
The  O(4) background admits a pair of zero modes with
opposite chirality.

\section{Solving the Dirac equation by inversion}

In this section we use the conformal transformation (\ref{Z5}) to map
the pair of O(4) zero mode states with turning point at $\xi=0$,
onto a pair of O(3)-symmetric solutions which
 solves the Dirac equation in the background LS configuration
described in section 2. They 
are localized (normalizable) in 3-space at all times. They describe
propagation of the produced quarks in 
Minkowski space-time into asymptotically free outgoing  chiral quarks,
with computable
spectra. We will also show that those solutions in fact  start with 
{\em zero energy}  at the turning time $t=0$: this  implies that the
sea contribution for positive and negative energy states is symmetric
and cancels in the chiral charge.

The solution itself is obtained by the following inversion formula
\be
{\bf Q}_+ (x) = \gamma_4\,\frac{\gamma_\mu \,(y +a)_\mu}{1/(y+a)^2}
\,{\bf \Psi}_+ (y)
\label{x1}
\ee
and it solves the ($\gamma_4\times$)
Dirac equation in the (Euclidean) LS gauge configuration (\ref{Z6x}).
Under the conformal transformation (\ref{Z5}) followed by $\gamma_4$
multiplication, the right-handed zero mode state (\ref{Z14}) remain
right-handed,
Unwinding $y$ in terms of $x$ and analytically continuing to Minkowski
space yield~\footnote{Note that we are still using the notation ${\bf C}$ for the
normalization of the zero energy state.}

\be
{\bf Q}_+ (t,r) = &&
\frac{{\bf C}}{((t+i\rho)^2-r^2)^2}\nonumber\\
&&\times \left((\rho-it) +i\vec{\sigma}\cdot\vec{x}\right)\,e^{-2F(\xi)}\,U_+\,\,,
\label{Z19}
\ee
with

\be
\xi=\frac 12 \,{\rm
ln}\left(\frac{(t-i\rho)^2-r^2}{(t+i\rho)^2-r^2}\right)\,\,.
\label{Z20}
\ee
The normalization is fixed at $t=0$ by noting that $F(0)=0$, i.e.

\be
{\bf Q}_+ (0,r) = 
\frac{{\bf C}}{(\rho^2+r^2)^2}
\,\left(\rho +i\vec{\sigma}\cdot\vec{x}\right)\,U_+\,\,,
\label{Z21}
\ee
and demanding that in 3-space,

\be
{2\,|{\bf C}|^2}\,\int\,d^3x\,\frac 1{(\rho^2+r^2)^3}=1\,\,,
\label{Z21x}
\ee
which gives $|{\bf C}|=\sqrt{2\rho^3}/\pi$.
We note that the norm is independent of which LS configuration
is chosen, making all mapped O(3) states normalizable. 

The next question we address is whether the solution for the exploding
sphaleron indeed starts with the
zero energy state.  Naively, at small t it does not look as a  static
solution at all. However, one can observe that the initial
time derivative of (\ref{Z19}) fulfills

\be
-i\dot{\bf Q}_+ (0,r)=\frac{(3-8\,f(0))
\,\rho-i\vec{\sigma}_s\cdot\vec{x}}{\rho^2+r^2}
\,{\bf Q}_+(0,r)\,\,.
\label{Z21xx}
\ee
This may be rewritten as a gauge phase

\be
-i\,\dot{\tilde{\bf Q}}_+ (0,r)= \dot{\bf\Lambda} (0,r) \,
\tilde{\bf Q}_+(0,r)\,\,,
\ee
with SU(N) and U(1) gauge phases. The former is
\be
{\rm ln}\,{\bf\Lambda}= -it\frac{\vec{\sigma}_s\cdot\vec{x}}{\rho^2+r^2}\,\,,
\ee
while the latter, $\tilde{\bf Q}_+=e^{it\Theta}{\bf Q}_+$, is

\be
\Theta = \frac{(3-8\,f(0))\,\rho}{\rho^2+r^2}\,\,.
\ee
Since $A_0=0$ at $t=0$,
the configurations (\ref{Z19}) have zero energy modulo an SU(2)$\times$U(1)
gauge transformation provided that $f(0)$ is real. The SU(2) is related
to the color gauge group, and the U(1) to an additional local symmetry
of the Dirac Hamiltonian ${\bf H}(t) =\gamma_4\vec\gamma\cdot\vec\nabla$
at $t=0$, namely

\be
e^{it\,\Theta}\,{\bf H} (t) \,e^{-it\,\Theta}
= {\bf H} (t) -i t\,\gamma_4\vec\gamma\cdot\vec\nabla\Theta
\ee
The solution thus looks static for the modified Hamiltonian, and the
difference between the transformed and original solution
is absent at $t=0$. We thus conclude that we {\it do} start from the zero
energy state, masked by  gauge transformations.  All the O(3) 
right-handed quarks at and above the sphaleron point are normalizable
zero energy states to the Minkowski Dirac equation with the LS background.

\begin{figure}[h]
\epsfxsize=6cm 
\epsfbox{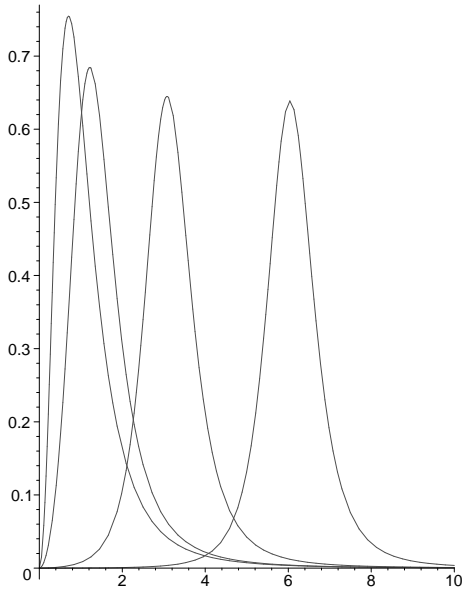}
\caption{\label{fig_density}
The radial density distribution $\varrho_+ (t,r)$ for
$t/\rho=0,1,3,6$: the set of curves starts at
the (anti)sphaleron point.}
\end{figure}

At large times $t\sim r\gg \rho$ with $v=r-t$,
(\ref{Z19}) simplifies to

\be
{\bf Q}_+ (t,r) = 
\frac {-i\,{\bf C}}{4t\,(v-i\rho)^2}\,
\left(1-\,\vec{\sigma}_s\cdot\hat{r}\right)\,e^{2F(\xi)\,\vec\sigma_s\cdot\hat{r}}\,U_+\,\,,
\label{Z22}
\ee
with 

\be
\xi=\frac 12 {\rm ln} \left( \frac{v+i\rho}{v-i\rho}\right)\,\,.
\label{Z23}
\ee
We have used the relation

\be
e^{-2F(\xi)}\left(1-\vec{\sigma}_s\cdot\hat{r}\right)
=\left(1-\vec{\sigma}_s\cdot\hat{r}\right)\,e^{2F(\xi)\,\vec{\sigma}_s\cdot\hat{r}}\,\,,
\label{Z23x}
\ee
to display the exponential as an SU(2) hedgehog gauge rotation, since
$F(\xi)$ is purely imaginary asymptotically.
The quark field is localized and normalized at $t=0$ and weaken
asymptotically as $1/t$ up to a gauge transformation. This is
a hallmark of a radiation field which admits a normal mode decomposition
as we will show below. The normalization is preserved by the
time-dependent evolution since

\be
\int\,d^3x\,|{\bf Q}_+|^2 (t, r) =\pi\,|{\bf C}|^2\,
\int_{-\infty}^{+\infty}\frac{dv}{(v^2+\rho^2)^2} = 1\,\,.
\label{Z23xx}
\ee

The radial density is then given by

\be
\varrho_+ (t, r) =4\pi\, r^2\, {\rm Tr}\left({\bf Q}_+^\dagger{\bf Q}_+\right)(t,r)
\label{Z24}
\ee
that is

\be
\varrho_+ (t,r) =\frac {16}\pi\,e^{-4{\rm Re}\,F(\xi)}\,
\frac{\rho^3\,r^2((\rho^2+t^2)+r^2)}
{((t^2+\rho^2-r^2)^2+4r^2\rho^2)^2}\,\,,
\label{Z24x}
\ee
at the (anti)sphaleron point as a function of the evolution time $t/\rho=0,1,3,6$.
At the escape point $t=0$, the zero energy state is fully localized in
space. At large times $t> \rho$ the energy density rapidly obtains a
frozen form, and eventually the
final quarks move luminally as free waves. A sample of density
profiles is shown in Fig.\ref{fig_density}, where it is easily seen
that the  time needed to reach the asymptotic form is 
$t_*\sim \rho$.

\section{Spectral Analysis of the outgoing quarks}

\begin{figure}[h]
\epsfxsize=5cm 
\epsfbox{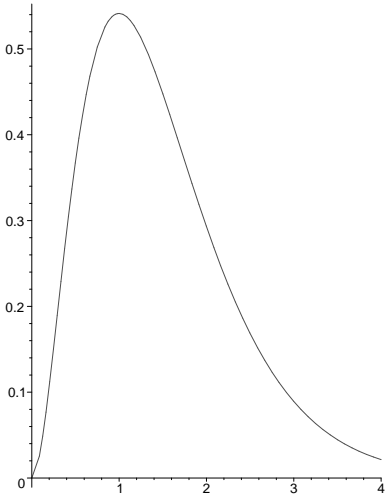}
\caption{\label{fig_spectrum}
The spectrum  of {\bf R} quarks released at the Sphaleron point
versus its momentum k in units of $1/\rho$.}
\end{figure}

A normal mode decomposition allows for a spectral analysis of the final
quark states released by the antisphaleron (sphaleron). Using the free
field decomposition asymptotically,

\be
&&{\bf Q}_+ (t,k) = \frac{(2\pi)^{\frac 32}}{\sqrt{2k}}
\,\left({\bf q}_R (\vec{k}) e^{-ikt} + {\bf q}_L^\dagger (-\vec k ) \,e^{+ikt}\right)\,\,,
\label{Z25}
\ee
allows the identification of the 1+1 right-mover ${\bf q}_R$
with the right-handed asymptotic chiral quark and the 1+1
left-mover ${\bf q}_L^\dagger$ with the left-handed asymptotic chiral 
anti-quark. In the antispaleron background we have

\be
&&\frac{\sqrt{2k}}{(2\pi)^{3/2}}{\bf Q}_+ (t,k) =\nonumber\\
&&
{\bf C}\sqrt{\pi\,k}\,\,\theta (k) \,e^{-k\rho + ikt} \,
(1-\vec{\sigma}_s\cdot\hat{k})\,U_+
\label{Z25x}
\ee
so that 

\be
{\bf q}_L^\dagger (k) = 2\rho^3\,\sqrt{\frac k{\pi}}\,e^{-k\rho}\,
(1-\vec{\sigma}_s\cdot\hat{k})\,U_+\,\,.
\label{Z25xx}
\ee
${\bf q}_R$ follows from (\ref{Z25xx}) by charge conjugation 
after mapping the O(4) conjugate zero mode to the O(3) zero
energy state,

\be
{\bf q}_R (k) = 2\rho^3\,\sqrt{\frac k{\pi}}\,e^{-k\rho}\,
(1+\vec{\sigma}_s\cdot\hat{k})\,U_-\,\,.
\label{Z25xxx}
\ee
The (chiral) density of left antiquarks and right quarks
are opposite asymptotically ${\bf n}_L=-{\bf n}_R$, with
\be
{\bf n}_R (k) = \frac {4\pi\,k^2}{2k}
 \,|{\bf q}_R^\dagger(\vec k )|^2 = \rho\,(2k\rho)^2\,e^{-2k\,\rho}\,\,.
\label{Z26}
\ee
Fig. \ref{fig_spectrum} shows the phase space distribution ${\bf n}_{R}/\rho$
of the right quarks released at the sphaleron point as a function of $k\rho$. 
The distribution integrates exactly to one produced quark,

\be
{\bf n}_R=\int_0^\infty \, dk\, {\bf n}_{R} (k) = 1\,\,.
\label{Z27}
\ee
The quark spectrum is close to Planckian with an effective
temperature $T=2/\rho$ of about 300 MeV for a standard $\rho=1/3$ fm.
Incidentally, this is close to the initial temperature of a quark-gluon
plasma in the RHIC energy domain.
The released quarks carry asymptotically a total energy

\be
{\bf M}_F=2\int_0^\infty\,dk\,k\,{\bf n}_R (k) = \frac{3}{\rho}\,\,,
\label{Z28}
\ee
which is small in the weak coupling limit under consideration, i.e.
${\bf M}_F/{\bf M}_S = \alpha/\pi\ll 1$. We recall that
in~\cite{jsz02} the number of prompt gluons released
was evaluated at $1.1/\alpha_s(\rho)$, which is parametrically large
in this limit. For this reason, back reaction of fermions onto gluons
can be neglected.

The situation is however quite different for `typical' QCD instantons
and sphalerons, with $\rho\sim 1/3$ fm.  In this case the produced
$2N_F=6$ quarks\footnote{We have ignored strange quark mass here,
because the corresponding scale at which the development occure is
$\sim 1/\rho \sim 1 \, GeV$.}  
should be compared to $\sim 3$ gluons  released 
classically from the explosion at the sphaleron
point. The total energy carried by one flavor species
would be $3/\rho\sim 1.8$ GeV, while the whole available energy (the
sphaleron mass) is only about 3 GeV.  
Needless to say that such results follow from a wrong assumption: 
in this regime the fermions are not a small perturbation
riding on the back of the gluonic wave. Their {\em back
reaction} on the YM fields cannot
be ignored. This imply a $joint$ solution of the coupled Yang-Mills
and Dirac equations
is needed, better yet
with vacuum polarization effects included. Although such
analysis is still missing, we may speculate that the spectra would
approximately look the same, with a rescaled energy per quark
appropriate for total energy conservation.

\section{Pair Production by Spectral Flow}

We have established that the release of an antisphaleron liberates
$N_F\,\overline{\bf L}{\bf R}$ pairs, while the release of a sphaleron
liberates $N_F\,\overline{\bf R}{\bf L}$ pairs. This mechanism operates
for all energies Q above and including the sphaleron point, and is
signaled by a pair of energy levels crossing at $t=0$ 
(pair production) in the form of zero energy states, with their
subsequent acceleration by pertinent gluoelectric fields.

Compare this to what happens at low energies, when the gauge field configurations
are well-separated instanton and anstiinstanton, and the production of
 $2 N_f$ fermions is described by the well-known 't
Hooft vertex. Although we have not shown it explicitly, we conjecture
that {\em the same} number of fermions is produced at any energy, by continuity.
 
On the other hand, the escaping gauge configurations above the sphaleron point were shown
to carry fractional Chern-Simons number~\cite{jsz02}

\be
{\bf N} (0) =\frac 12\,\left(\frac{Q}{M_S}\right)^{2/5}\,\,.
\label{1}
\ee
The {\em non-integer} character of (\ref{1}) calls for a key question:
{\em How does the conventional Adler-Bell-Jackiw (ABJ) anomaly
with integer fermion number work?}  
The answer is as follows: 
The net chirality in the Dirac spectrum is carried by both the 
valence and sea parts,

\be
\Delta{\bf n}(t) =&&({\bf n}_R-{\bf n}_L) (t)\nonumber\\ =&& \Delta{\bf n}_V (t) 
+\Delta{\bf n}_S(t)\,\,.
\label{PP1}
\ee
At t=0 the Dirac spectrum associated to 
${\bf H} (t)$
is C-conjugate in the O(3) LS configuration
except for the two zero energy states,

\be
\Delta{\bf n}(0) =\Delta{\bf n}_S (0)= 2N_F\,{\bf N} (0)\,\,.
\label{PP2}
\ee
For $t>0$, one level dives in and one-level pulls out.
The chirality count changes discretely by 2.
In particular,
\be
&&\Delta{\bf n}_V (t)=2N_F\nonumber\\
&&\Delta{\bf n}_S (t)=2N_F({\bf N}(t)-1)
\label{PP3}
\ee
wherein the valence part is t-independent. Asymptotically, the
chiral polarization in the Dirac spectrum carried by the sea
quarks vanishes $\Delta{\bf n}_S(\infty)=0$, since the O(3) asymptotic
configuration is commensurate with plane waves, thus C-conjugate. The net result
is

\be
\Delta{\bf n} (\infty)= \Delta{\bf n}_V (\infty) =2N_F\,\,,
\label{PP4}
\ee
which is the result explicitly obtained from the zero energy states
released.


Some unsatisfied readers may ask how one can generalize index
theorems, so that the number of level crossing would be calculable
directly from the gauge field itself, without the explicit solution of 
the Dirac eqn. Or, in a more practical form of the same,
Does the conformal mapping
exhausts $all$ level crossings of the time-dependent Hamiltonian
${\bf H} (t)$ for all $Q\geq M_S$ ?

Interesting work toward answering these questions can be found in
recent work~\cite{KL}, where one can also find earlier references.
 In particular, it was found that
 the answer to the last one is negative.
In a numerical analysis presented there
it was found that (in our units)  for $Q>19.45\,M_S$ other crossings
 in the LS background do occur. (However, 
 these crossings are for practical purposes irrelevant since the cross
section
there is clearly too small.)

To summarize: Sphaleron (antisphaleron) production leads to pair production of
chiral quarks by spectral flow, and it produces one pair of quarks for 
each light flavor with unit probablity. 
This  nonperturbative mechanism for prompt chiral production
is not
suppressed by powers of $\alpha_s$ as in pQCD, or
exponentially
hampered by the constituent quark masses as in the string production mechanism. 
The qualitative effects of the current quark masses is to cause a power
suppression in the production mechanism as opposed to the exponential
suppression in the string.

\section{Conclusions}

We have started with zero mode solutions to the
O(4) configurations in Euclidean space that solves exactly the
Yang-Mills equations. We then have shown that through a pertinent 
conformal mapping and analytical continuation they map exactly on
the spherically O(3)-symmetric LS solutions \cite{LS} with a turning point at $t=0$ describing
explosion of the YM sphaleron-like clusters \cite{OCS}. We then have used
the same conformal mapping to construct a class of normalizable
solutions to the Dirac equation, in the LS background. We have
shown that these solutions do correspond to the zero energy states at $t=0$ for 
all LS configurations with energies larger or equal to the 
sphaleron mass. 

By following their evolution in time, we have explicitly shown that
the zero energy states becomes at large time free Dirac outgoing waves with 
fixed chirality $\pm 1$ and exactly 1 quantum. Thus sphaleron
(antisphalerons) production
 liberates $2N_F$ chiral quarks. We have
identified this liberation by a pair level crossing at
$t=0$ in the Dirac spectrum. We have used the asymptotic
zero energy states to construct the pertinent quark spectra
 produced. 

Many more things can be done based on the results reported here. One
is an account of back reaction of fermions on the YM fields, as
discussed in the previous section. Another is a 
projection of the outgoing waves to the wave function of outgoing
hadrons, with many detailed predictions about exclusive and inclusive
hadronic production to follow as we will report elsewhere.

The present results are of interest to both hadron-hadron
and nucleus-nucleus collisions at large $\sqrt{s}$. Simple
arguments indicate that while in hadron-hadron scattering production
via sphalerons, it is only a small part of the multiplicity, in
 nucleus-nucleus collisions it is not so and hundreds of
sphalerons may be released, contributing substantially to
the total entropy  produced~\cite{ES_how}. In view of that, 
the production of $2N_F$ chiral quarks per cluster may make the
quark-gluon plasma produced in heavy-ion collisions quark rich, contrary 
to expectations based on perturbation theory or the color glass
description~\cite{LR}.


\vskip 1.25cm
{\bf Acknowledgments}
\\\\
This work was supported in parts by the US-DOE grant
DE-FG-88ER40388.



\begin{thebibliography}{99}
\bibitem{electroweak}
A.~Ringwald, Nucl.Phys. {\bf B330} (1990) 1, 
O.~Espinosa, Nucl.Phys. {\bf B343} (1990) 310;
V.~Khoze, A. Ringwald, Phys. Lett. {\bf B259} (1991) 106.

\bibitem{Manton}
N.~Manton, Phys.Rev. {\bf D 28} (1983) 2019; 
F.~Klinkhamer and N.~Manton, Phys. Rev. {\bf D30} (1984) 2212.

\bibitem{sphaleron_decay}
J.~Zadrozny,
 Phys. Lett. {\bf B284} (1992) 88;
 M.~Hellmund and J.~Kripfganz,
 Nucl. Phys. {\bf B373} (1992) 749.


\bibitem{sz01}
E.~Shuryak and I.~Zahed, Phys. Rev. {\bf D62} (2000) 085014; 
M.~Nowak, E.~Shuryak and I.~Zahed, Phys. Rev. {\bf D64} (2001) 034008.

\bibitem{KKL}
D.~Kharzeev,~Y.~Kovchegov~and~E.~Levin, Nucl. Phys. {\bf A690} (2001) 621.

\bibitem{shifman}
V.~Zakharov, Nucl. Phys. {\bf B353} (1991) 683;
M.~Maggiore and M.~Shifman, Phys. Rev. {\bf D46} (1992) 3550.

\bibitem{khlebnikov}
S.~Khlebnikov, Phys. Lett. {\bf B282} (1992) 459.

\bibitem{DP94}
D.~Diakonov and V.~Petrov, Phys. Rev. {\bf D 50} (1994) 266.


\bibitem{jsz02}
R.~Janik, E.~Shuryak and I.~Zahed, {\tt hep-ph/0206005}.


\bibitem{LS}
M.~Luscher, Phys. Lett. {\bf B70} (1977) 321;
B.~Schechter, Phys. Rev. {\bf D16} (1977) 3015.

\bibitem{OCS}
D.~Ostrovsky, G.~Carter and E.~Shuryak,
{\tt hep-ph/0204224}.


\bibitem{phi4}
V.~Rubakov, D.~Son and P.~Tinyakov, Nucl. Phys. {\bf B404} (1993) 65;
D.~Son and P.~Tinyakov, Nucl. Phys. {\bf B415} (1994) 101.

\bibitem{book}
M.~Rho, M.~Nowak and I.~Zahed, `Chiral Nuclear Dynamics', World
Scientific; p. 42.


\bibitem{KL}
F.~Klinkhamer and Y.~Lee, Phys. Rev. {\bf D64} (2001) 065024.


\bibitem{LR}
L.~McLerran~and~R.~Venugopalan, Phys. Rev. {\bf D49} (1994) 2233; 3352.

\bibitem{ES_how}
E.~Shuryak, {\tt hep-ph/0205031}.

\end{thebibliography}
\end{document}